\documentclass[amsmath, amssymb, aps,prl, superscriptaddress, preprint]{revtex4-1}

\usepackage[dvipdfmx]{graphicx}
\usepackage{bm}
\usepackage[top=20truemm,bottom=20truemm,left=20truemm,right=20truemm]{geometry}
\usepackage{array}
\usepackage{color}

\begin{document}


\title{Electronic Structure of the Ferromagnetic Semiconductor Fe-doped Ge Revealed by Soft X-ray Angle-Resolved Photoemission Spectroscopy}


\author{S. Sakamoto}
\affiliation{Department of Physics, The University of Tokyo, Bunkyo-ku, Tokyo 113-0033, Japan}

\author{Y. K. Wakabayashi}
\affiliation{Department of Electrical Engineering and Information Systems, The University of Tokyo, Bunkyo-ku, Tokyo 113-8656, Japan}

\author{Y. Takeda}
\affiliation{Materials Sciences Research Center, Japan Atomic Energy Agency (JAEA), Sayo-gun, Hyogo 679-5148, Japan}

\author{S.-i. Fujimori}
\affiliation{Materials Sciences Research Center, Japan Atomic Energy Agency (JAEA), Sayo-gun, Hyogo 679-5148, Japan}

\author{H. Suzuki}
\affiliation{Department of Physics, The University of Tokyo, Bunkyo-ku, Tokyo 113-0033, Japan}

\author{Y. Ban}
\affiliation{Department of Electrical Engineering and Information Systems, The University of Tokyo, Bunkyo-ku, Tokyo 113-8656, Japan}

\author{H. Yamagami}
\affiliation{Materials Sciences Research Center, Japan Atomic Energy Agency (JAEA), Sayo-gun, Hyogo 679-5148, Japan}
\affiliation{Department of Physics, Kyoto Sangyo University, Kyoto 603-8555, Japan}

\author{M. Tanaka}
\affiliation{Department of Electrical Engineering and Information Systems, The University of Tokyo, Bunkyo-ku, Tokyo 113-8656, Japan}
\affiliation{Center for Spintronics Research Network, The University of Tokyo,  Bunkyo-ku, Tokyo 113-8656, Japan}

\author{S. Ohya}
\affiliation{Department of Electrical Engineering and Information Systems, The University of Tokyo, Bunkyo-ku, Tokyo 113-8656, Japan}
\affiliation{Center for Spintronics Research Network, The University of Tokyo,  Bunkyo-ku, Tokyo 113-8656, Japan}

\author{A. Fujimori}
\affiliation{Department of Physics, The University of Tokyo, Bunkyo-ku, Tokyo 113-0033, Japan}


\date{\today}

\begin{abstract}
Ge$_{1-x}$Fe$_{x}$ (Ge:Fe) shows ferromagnetic behavior up to a relatively high temperature of 210 K, and hence is a promising material for spintronic applications compatible with Si technology.
We have studied its electronic structure by soft x-ray angle-resolved photoemission spectroscopy (SX-ARPES) measurements in order to elucidate the mechanism of the ferromagnetism. We observed finite Fe 3$d$ components in the states at the Fermi level ($E_{\rm F}$) in a wide region in momentum space and $E_{\rm F}$ was located above the valence-band maximum (VBM).
First-principles supercell calculation also suggested that the $E_{\rm F}$ is located above the VBM, within the narrow spin-down $d$($e$) band and within the spin-up impurity band of the deep acceptor-level origin derived from the strong $p$-$d$($t_{2}$) hybridization. We conclude that the narrow $d$($e$) band is responsible for the ferromagnetic coupling between Fe atoms while the acceptor-level-originated band is responsible for the transport properties of Ge:Fe.
\end{abstract}


\maketitle

Ferromagnetic semiconductors (FMSs) such as (Ga,Mn)As \cite{Ohno:1996aa, Ohno:1998aa} have attracted much attention both from scientific and technological points of view \cite{Wolf:2001aa, Igor, dietl2010ten, Dietl:2014aa, Jungwirth:2014aa, Tanaka:2014aa}. 
Group-IV FMSs are particularly important because they are compatible with mature Si-based technology. 
Ge$_{1-x}$Fe$_{x}$ (Ge:Fe) is a promising material \cite{Shuto:2006aa, Shuto:2007aa, Wakabayashi:2014aa, Wakabayashi:2014ab}, and indeed can be grown epitaxially on Ge and Si substrates by the low-temperature molecular beam epitaxy (LT-MBE) method without the formation of intermetallic precipitates \cite{Ban:2014aa}. 
It shows $p$-type conduction, but the carrier concentration of $\sim$10$^{18}$ cm$^{-3}$ \cite{Ban:2014aa} is orders of magnitude smaller than that of doped Fe atoms ($\sim$10$^{21}$ cm$^{-3}$). The Curie temperature ($T_{\rm C}$) increases with the Fe content and with the inhomogeneity of Fe atom distribution \cite{Wakabayashi:2014aa, Wakabayashi:2014ab}, and reaches $\sim$210 K at highest by post-growth annealing \cite{Wakabayashi:2014aa}, which is above the highest $T_{\rm C}$ of (Ga,Mn)As, $\sim$200 K \cite{Chen:2011aa}. Unlike (Ga,Mn)As, the $T_{\rm C}$ does not depend on carrier concentration \cite{Ban:2014aa}. 
Recent x-ray absorption spectroscopy (XAS) and x-ray magnetic circular dichroism (XMCD) measurements \cite{Wakabayashi:2016aa} have revealed the valence of Fe substituting Ge to be 2+, which indicates that each Fe atoms would provide two holes.
It was also found that nanoscale ferromagnetic domains exist even above the $T_{\rm C}$, the origin of which was attributed to the inhomogeneous distribution of Fe atoms on the nanoscale. 

In order to explain the origin of the ferromagnetism in (Ga,Mn)As and related FMSs, two models have been proposed so far \cite{Jungwirth:2006aa, Sato:2010aa, dietl2010ten}, namely, the valence-band model \cite{Dietl:2000aa, Dietl:2001aa} and the impurity-band model \cite{Okabayashi:2001aa, Sato:2002aa, Burch:2006aa, ohya2011nearly}.
In the valence-band model, acceptor levels derived from the magnetic impurities are merged into the valence band and itinerant holes occupying states around the valence-band maximum (VBM) mediate ferromagnetism through Zener's $p$-$d$ exchange mechanism. 
In the case of the impurity-band model, on the other hand, impurity levels are detached from the VBM and lies within the band gap of the host semiconductor and hence ferromagnetism is stabilized through a double-exchange-like mechanism within the impurity band. 
In order to elucidate the electronic structure of Ge:Fe, especially, the position of the Fermi level ($E_{\rm F}$) and the modification of the host band structure caused by the Fe 3$d$ electrons, we have performed soft x-ray angle-resolved photoemission spectroscopy (SX-ARPES) measurements and  first-principles supercell calculations.

\begin{figure}
\begin{center}
\includegraphics[width=8 cm]{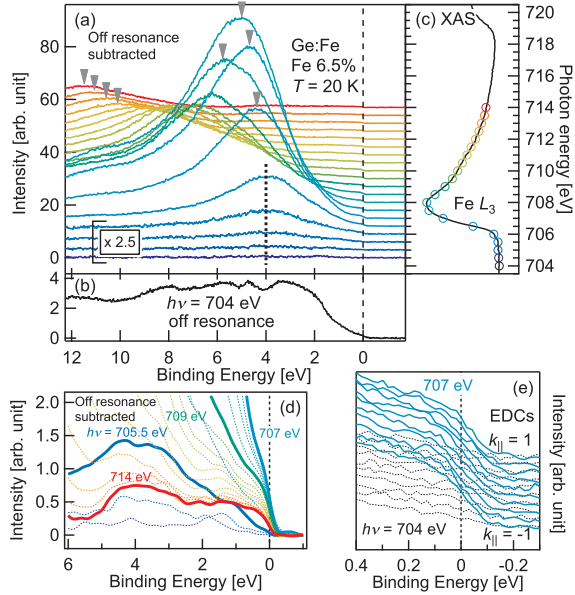}
\caption{Resonance photoemission spectra of Ge$_{0.9335}$Fe$_{0.065}$. (a) Spectra taken in the angle-integrated mode across the Fe $L_{3}$ absorption edge at 0.5 eV photon-energy interval as depicted by circles on the XAS spectrum in panel (c). The color of the open circles in panel (c) corresponds to that of the spectra in panel (a). The off-resonance spectrum shown in panel (b) has been subtracted from all the spectra in panel (a), where and the units of the vertical axes in panels (a) and (b) are the same. Triangles show the position of the normal Auger peak. The spectra for $h\nu=$ 704.5 - 706 eV have been magnified by a factor of 2.5. (d) Enlarged plot of the spectra in panel (a). The same color as in panel (a) is used. (e) Energy distribution curves taken in the angle-resolved mode along $k_{\parallel}$ $\parallel$ [-110], where $k_{\parallel}$ is in units of $2\sqrt{2}\pi$/$a$.}
\label{AIPES}
\end{center}
\end{figure}

A Ge$_{0.935}$Fe$_{0.065}$ film was synthesized using the LT-MBE method at the growth temperature of 240 $^{\circ}$C. The structure of the sample was, from the top surface to the bottom, Ge cap ($\sim$ 2 nm)/Ge$_{0.935}$Fe$_{0.065}$ ($\sim$130 nm)/Ge buffer ($\sim$20 nm)/p-Ge (001) substrate. The crystal orientation of the Ge:Fe sample was confirmed to coincide with that of the Ge substrate.
The $T_{\rm C}$ was estimated to be 100 K from the growth condition. 
In order to remove the oxidized surface layer, just before loading the sample into the vacuum chamber of the spectrometer, we etched the sample in a hydrofluoric acid (HF) solution (3 mol/L) for 5 seconds and subsequently rinsed it in water, which is known to be an efficient way to clean the surfaces of Ge \cite{Sun:2006aa} as well as those of Ge:Fe \cite{Wakabayashi:2016aa}. 

SX-ARPES experiment was performed at beam line BL23SU of SPring-8. The sample temperature was set to 20 K and circular polarized x rays of 700-950 eV were used.
The energy resolution was about 170 meV. The sample was placed so that the [-110] direction became parallel to the analyzer slit and perpendicular to the beam. By rotating the sample around the [-110] axis and changing the photon energy, we could cover the entire Brillouin zone. X-ray absorption spectra were taken in the total electron yield mode. 

First-principles supercell calculations were done based on the density functional theory (DFT) utilizing the full-potential augmented-plane-wave method implemented in the WIEN2k package \cite{blaha2001wien2k}. 
For the calculation of the host Ge band structure, modified Becke-Johnson (mBJ) exchange potential with the local density approximation (LDA) for correlation potential \cite{tran:2009aa} was employed. 
For the calculation of the spin-resolved partial density of states (PDOS) of Fe 3$d$ in Ge, we constructed a 3$\times$3$\times$3 supercell consisting of 53 Ge and one Fe atoms, and used the generalized gradient approximation (GGA) of Perdew-Burke-Ernzerhof type \cite{perdew:1996aa} for the exchange-correlation energy functional.
The experimental lattice constant of $a=5.648$\ \AA\ for Ge$_{0.935}$Fe$_{0.065}$ \cite{Wakabayashi:2014ab} was used and spin-orbit interaction was included for both calculations.


Figure \ref{AIPES}(a) shows resonance photoemission (RPES) spectra taken in the angle-integrated mode at 0.5 eV photon-energy intervals in the Fe $L_{3}$ absorption-edge region. 
Here, the off-resonance spectrum taken at a lower photon energy of 704 eV has been subtracted. The color of the spectra correspond to that of the open circles on the XAS spectra in Fig. \ref{AIPES}(c) and indicate photon energies. Note that the binding energy is defined relative to $E_{\rm F}$.
One can see a strong normal Auger peak dispersing with photon energy in the spectra. This indicates the itinerant nature of the Fe 3$d$ electrons in Ge:Fe, because the normal Auger process takes place when the core-hole potential is screened by conduction electrons faster than core-hole decay. 
The itinerant nature of the Fe 3$d$ electrons is further confirmed by the XAS spectra consisting of a broad single peak without multiplet structure seen when 3$d$ electrons are localized \cite{Laan:1992aa}. 
It should be noted that the XAS spectrum does not show Fe$^{3+}$ oxides signals, which guarantees the effectiveness of the HF etching prior to the measurements. 
In addition to the normal Auger peak, non-dispersive feature can be seen around the binding energy of 4 eV denoted by a dashed line, and exhibits resonance enhancement. (How the dispersive and non-dispersive features coexist in the spectra are summarized in Fig. S1 \cite{SM}.) 
Such a structure with a constant binding energy is either due to direct recombination, where the photoexcited electron recombines with the core hole, or to a satellite \cite{Thuler:1982aa}, where the photoexcited core electron acts as a spectator to the core-hole recombination process.

Figure \ref{AIPES}(d) shows the same RPES spectra plotted on an expanded scale. 
Due to the strong Auger peak, it was difficult to extract the PDOS from the spectra taken with the photon energy of the absorption peak at 708 eV. Therefore, by using a higher energy photons of 714 eV, we have deduced the Fe 3$d$ PDOS as shown by a red curve in Fig. \ref{AIPES}(d). 
The PDOS is broad extending from $E_{\rm F}$ to 5 eV below it, out of which the structure around 4 eV is attributed to a satellite because it showed strong enhancement at the resonance energy like the satellite in transition metals and transition-metal compounds listed in Supplementary Material \cite{SM}.
Therefore, we consider that the main part of the Fe 3$d$ PDOS is located from $E_{\rm F}$ to $\sim$3 eV below it.  
In addition, there can be seen the Fermi edge-like step at $E_{\rm F}$, which indicates that the Fe 3$d$ states have a finite contribution to the states at $E_{\rm F}$, and are involved in the charge transport of Ge:Fe. 
Figure \ref{AIPES}(e) shows the energy distribution curves (EDCs) taken in the angle-resolved mode at the photon energies of 704 eV (off-resonance) and 707 eV (on-resonance). The enhanced Fe 3$d$ states were found to exist in a wide region in momentum space without appreciable dispersions.
Note that the Fermi edge-like feature at $E_{\rm F}$ is much clearer in Ge:Fe than in (Ga,Mn)As \cite{kobayashi:2014aa}, indicating that contributions of 3$d$ electrons to states at $E_{\rm F}$ are more pronounced in Ge:Fe than in (Ga,Mn)As.

\begin{figure}
\begin{center}
\includegraphics[width=8 cm]{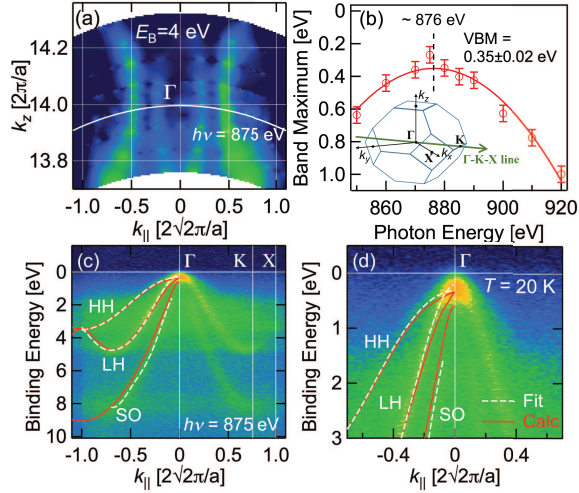}
\caption{ARPES band mapping for Ge$_{0.935}$Fe$_{0.065}$. (a) $k_{\parallel}$-$k_{z}$ mapping image at the binding energy of 4 eV. A white curve represents the ARPES cut for the photon energy of 875 eV. (b) Maximum of the band dispersion along $k_{\parallel}$ as a function of photon energy. The solid curve represents a fitted parabolic function. Inset shows the Brillouin zone of the fcc lattice. (c), (d) ARPES spectra along the $\Gamma$-K-X line taken with $h\nu = 875$ eV. The peak positions of the second derivatives of the EDCs have been fitted to a Fourier series and are shown by dashed curves. Solid curves represent the calculated band dispersions of the host Ge, where the heavy-hole (HH) band, the light-hole (LH) band, and the split-off (SO) band can be seen.}
\label{ARPES}
\end{center}
\end{figure}

Figure \ref{ARPES}(a) shows the photon energy dependence of ARPES spectra at the binding energy of 4 eV around the $\Gamma$ point, from which one can see that the ARPES taken with x rays of 875 eV crosses the $\Gamma$ point. From this plot using the final free-electron final-state model \cite{Liebowitz:1978aa},  the inner potential was determined to be 11 eV. 
In Fig. \ref{ARPES}(b), the maximum energy of the valence-band dispersion is plotted against photon energy, and reaches the VBM at $\sim876$ eV. The energy of the VBM thus deduced is found to be 0.35 eV below $E_{\rm F}$, indicating that the Fermi level of Ge:Fe is located in the middle of the Ge band gap of $\sim$0.7 eV.

Figures \ref{ARPES}(c) and \ref{ARPES}(d) show ARPES spectra along the ${\rm \Gamma}$-K-X line in the Brillouin zone of the fcc lattice (see the inset of Fig. \ref{ARPES}(b)) taken with the photon energy of 875 eV. 
The peak positions of the second derivatives of the EDCs have been fitted to a Fourier series and shown by dashed curves. Here, clear band dispersions characteristic of Ge, such as the heavy-hole (HH) band, the light-hole (LH) band, and the split-off (SO) band, can be seen, which indicates the good crystallinity of the Ge:Fe sample as well as the good quality of the sample surface after the HF etching. Solid curves represent the calculated band dispersions of the Ge host. 
As can be seen from Fig. \ref{ARPES}, the ARPES spectra of Ge:Fe agree fairy well with the calculated band dispersions of Ge, indicating that the doped Fe atoms did not affect the electronic structure of the Ge host significantly. Note that this is also the case for (Ga,Mn)As \cite{kobayashi:2014aa}.


\begin{figure}
\begin{center}
\includegraphics[width=8 cm]{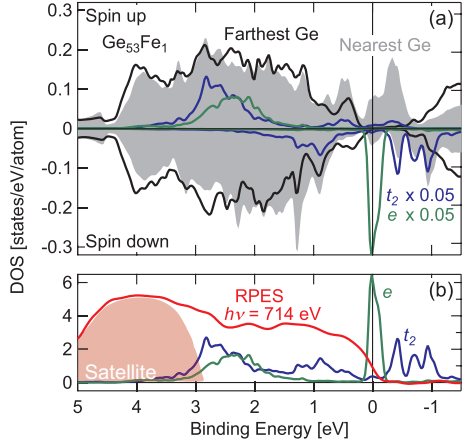}
\caption{Partial densities of states (PDOSs) of a $3\times3\times3$ supercell containing 53 Ge and one Fe atoms, corresponding to Ge$_{1-x}$Fe$_{x}$ ($x\sim1.85\%$). (a) Spin-resolved density of states. Black curve and gray area represent the PDOS of the farthest and the nearest Ge atom to the Fe atom, respectively, and blue and green curves represent the PDOS of the Fe 3$d$($t_{2}$) and 3$d$($e$) states of Fe, respectively. Here, the PDOS of the $t_{2}$ and $e$ states have been scaled by a factor of 0.05 for the sake of comparison with the PDOS of Ge. (b) Spin-averaged PDOS of the Fe 3$d$($t_{2}$) and 3$d$($e$) states of Fe 3$d$. The experimental spectrum is superposed by a red curve.}
\label{DFT}
\end{center}
\end{figure}

\begin{figure*}
\begin{center}
\includegraphics[width=16 cm]{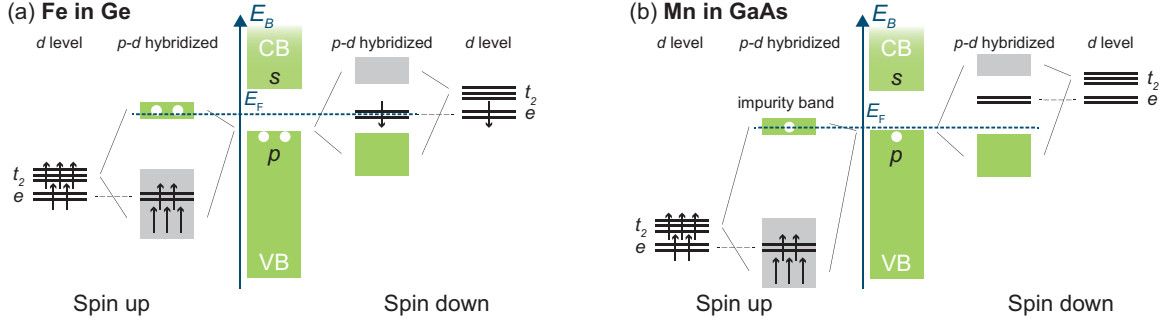}
\caption{Schematic energy-level diagram of (a) Ge:Fe and (b) (Ga,Mn)As. At the center of each panel, the valence band (VB) and the conduction band (CB) of the host semiconductor are shown. At the left- and right-hand sides of each panel, the spin-up and spin-down $d$ levels are shown, respectively. In addition, energy levels with $p$-$d$ hybridization are shown in between, where green and gray boxes represent the state with predominant $p$ and $t_{2}$ character, respectively.}
\label{Schematic}
\end{center}
\end{figure*}

In order to examine the electronic structure of a Fe atom substituting a Ge atom in the Ge host in comparison with a Mn atom substituting Ga in the GaAs host, we have calculated the spin-resolved DOS of a $3\times3\times3$ supercell containing 53 Ge atoms and one Fe atom substituting for Ge, corresponding to Fe $\sim$1.85\%-doped Ge as shown in Fig. \ref{DFT}(a).
Black curve and gray area represent the PDOS of the farthest and the nearest Ge atoms to the Fe atom, and blue and green curves represent the PDOS of Fe 3$d$($t_{2}$) and 3$d$($e$) orbitals, respectively.
The PDOS of the farthest Ge is not affected by the presence of Fe significantly, which means that the Fe atom in this supercell can be considered as an isolated impurity. 
On the other hand, the PDOS of the nearest Ge is strongly affected by hybridization with Fe 3$d$ states (mainly with Fe 3$d$($t_{2}$) states), in particular within $\sim$0.5 eV of $E_{\rm F}$, as in the case of (Ga,Mn)As. A significant difference between Ge:Fe and (Ga,Mn)As is that there is an additional Fe 3$d$ electron in Ge:Fe which occupies the minority-spin 3$d$($e$) states at the Fermi level. 
This means that Fe is in the Fe$^{2+}$ state with 3$d^{6}(sp)^{2}$ configuration, consistent with the XAS and XMCD measurements \cite{Wakabayashi:2016aa} and a previous calculation on a $2 \times 2 \times 2$ Ge supercell having a neighboring Fe-Fe pair \cite{Weng:2005aa}. (In that calculation, the 3$d$($e$) state was split into bonding and anti-bonding states due to the overlap of the $d$ orbitals of paired Fe atoms.)
Such an electronic structure was also found in the LDA calculations on (Ga,Fe)As \cite{Sandratskii:2003aa} and (In,Fe)As:Be \cite{Vu:2014aa}. In addition, the $p$-$d$($t_{2}$) hybridized states in Ge:Fe is pushed from the VBM into the band gap of host Ge and act as deep acceptor levels. The $E_{\rm F}$ appears to be located $\sim$0.2 eV above the VBM of the farthest Ge. The value of $\sim$0.2 eV is smaller than the experimental value of 0.35 eV. This is probably due to the existence of $\sim$15\% of interstitial Fe atoms \cite{Wakabayashi:2014ab}, which provide two electrons per Fe atom to the $sp$ orbitals and partially compensates holes.

Figure \ref{DFT}(b) shows the spin-averaged PDOS of Fe 3$d$($t_{2}$) and 3$d$($e$) orbitals in comparison with the experimentally obtained PDOS. Except for the structure around 4 eV, which we attribute to a satellite, the calculated PDOS agrees well with the experiment at least qualitatively, that is, both PDOS have a finite value at $E_{\rm F}$ and extend down to $\sim$3 eV below $E_{F}$.

A schematic energy-level diagram of the electronic structure of the Fe atom in the Ge matrix thus obtained is shown in Fig. \ref{Schematic}(a) and that of the Mn atom in the GaAs matrix in Fig. \ref{Schematic}(b).
In both cases, due to the $T_{d}$ local crystal symmetry around the transition-metal atom, the $d$ levels are split into two sublevels, the doubly degenerate 3$d$($e$) level and the triply degenerate 3$d$($t_{2}$) level. 
In the presence of $p$-$d$ hybridization (predominantly $p$-$d$($t_{2}$) hybridization), the spin-up 3$d$($t_{2}$) levels are shifted downwards and the spin-down $t_{2}$ levels upwards. At the same time, some $p$ states are split from the VBM: spin-up levels are shifted upward and spin-down ones downward. 
Note that, as a result of the $p$-$d$($t_{2}$) hybridization, the shifted levels have both $d$($t_{2}$) and $p$ characters and, therefore, we refer to the lower levels as bonding levels, and the upper as antibonding levels hereafter.
In the case of (Ga,Mn)As, the spin-up $d$ levels are fully occupied and the spin-down $d$ levels are empty. Mn takes the Mn$^{\rm 2+}$ state with five spin-up $d$ electrons and one $p$ hole enters the valence band. 
Due to the strongest Hund's coupling of the Mn$^{2+}$ ion with $d^{5}$ configuration, the spin-up $d$ levels are located well below $E_{\rm F}$, while the spin-down $d$ levels are located well above $E_{\rm F}$. 
Therefore, the hole enters the spin-up antibonding levels with predominant $p$ characters split-off from the VBM and acts as a shallow acceptor level. 
On the other hand, from the electron counting argument \cite{SM}, the Fe atom substituting Ge should have six $d$ electrons and provides two $p$ holes. The spin-up $d$ levels of Ge:Fe are shallower in energy than those of (Ga,Mn)As because of the reduced Hund's energy, and $p$-$d$($t_{2}$) hybridization becomes stronger.
As a result, the spin-up antibonding levels are pushed well above the VBM compared to the Mn case and even above the spin-down 3$d$($e$) level. 
Therefore, the sixth $d$ electrons of Fe occupy the spin-down 3$d$($e$) states and the two $p$ holes reside in the spin-up states of the deep acceptor-level origin.
If the Fe concentration is high enough and Fe-Fe interaction is non-negligible, the band width of the spin-down 3$d$($e$) band would become broader, and the double-exchange mechanism would become effective.

From the above considerations, we conclude that the valence-band model or mean-field $p$-$d$ Zener model is not applicable in a different sense from the (Ga,Mn)As case.
The spin-up $p$-$d$($t_{2}$) hybridized levels located above the VBM appear responsible for the charge transport and the non-dispersive Fe 3$d$ intensity at $E_{\rm F}$ observed by the resonance ARPES measurements.
On the other hand, the narrow-band or nearly localized Fe 3$d$($e$) electrons play an essential role in stabilizing the ferromagnetism most likely through a double-exchange-like mechanism between neighboring Fe atoms.
The present picture explains the observed increase of $T_{\rm C}$ with Fe concentration \cite{Shuto:2006aa} and with the inhomogeneity of Fe distribution \cite{Wakabayashi:2014aa}. The same picture explains the observation of nanoscale ferromagnetic domains formed in Fe-rich regions well above the $T_{\rm C}$ \cite{Wakabayashi:2016aa}.

In summary, we have performed SX-ARPES measurements on Ge$_{0.935}$Fe$_{0.065}$. 
In the resonance photoemission spectra, a strong normal Auger peak could be seen, indicating the itinerant nature of the Fe 3$d$ electrons. 
ARPES spectra show that the Fermi level is located at 0.35 eV above the VBM and that non-dispersive Fe $3d$ states exist at the Fermi level, which can be attributed to spin-up $p$-$d$($t_{2}$) antibonding states of deep acceptor-level origin, and also to spin-down Fe 3$d$($e$) states.
Combining the ARPES result with the results of supercell calculations and the previous XMCD study, it is concluded that the ferromagnetic interaction is mediated by double-exchange interaction within the nearly localized down-spin Fe 3$d$($e$) band, and that charge transport occurs through the spin-up impurity band of the deep acceptor-level origin. 

This work was supported by Grants-in-Aid for Scientific Research from the JSPS (No. 15H02109, No. 23000010, and No. 26249039).
The experiment was done under the Shared Use Program of JAEA Facilities (Proposal No. 2014B-E29) with the approval of the Nanotechnology Platform Project supported by MEXT.
The synchrotron radiation experiments were performed at the JAEA beamline BL23SU in SPring-8 (Proposal No. 2014B3881).
A.F. is an adjunct member of Center for Spintronics Research Network (CSRN), the University of Tokyo, under Spintronics Research Network of Japan (Spin-RNJ).
S.S. acknowledges financial support from Advanced Leading Graduate Course for Photon Science (ALPS), and Y. K. W. acknowledges financial support from Materials Education program for the future leaders in Research, Industry, and Technology (MERIT). 
H.S. and Y.K.W. acknowledge financial support from JSPS Research Fellowship for Young Scientists.


\bibliography{BibTex_all_GeFe.bib}
\end{document}



\title{Electronic Structure of the Ferromagnetic Semiconductor Fe-doped Ge Revealed by Soft X-ray Angle-Resolved Photoemission Spectroscopy \\(Supplementary Information)}


\author{S. Sakamoto}
\affiliation{Department of Physics, The University of Tokyo, Bunkyo-ku, Tokyo 113-0033, Japan}

\author{Y. K. Wakabayashi}
\affiliation{Department of Electrical Engineering and Information Systems, The University of Tokyo, Bunkyo-ku, Tokyo 113-8656, Japan}

\author{Y. Takeda}
\affiliation{Materials Sciences Research Center, Japan Atomic Energy Agency (JAEA), Sayo-gun, Hyogo 679-5148, Japan}

\author{S.-i. Fujimori}
\affiliation{Materials Sciences Research Center, Japan Atomic Energy Agency (JAEA), Sayo-gun, Hyogo 679-5148, Japan}

\author{H. Suzuki}
\affiliation{Department of Physics, The University of Tokyo, Bunkyo-ku, Tokyo 113-0033, Japan}

\author{Y. Ban}
\affiliation{Department of Electrical Engineering and Information Systems, The University of Tokyo, Bunkyo-ku, Tokyo 113-8656, Japan}

\author{H. Yamagami}
\affiliation{Materials Sciences Research Center, Japan Atomic Energy Agency (JAEA), Sayo-gun, Hyogo 679-5148, Japan}
\affiliation{Department of Physics, Kyoto Sangyo University, Kyoto 603-8555, Japan}

\author{M. Tanaka}
\affiliation{Department of Electrical Engineering and Information Systems, The University of Tokyo, Bunkyo-ku, Tokyo 113-8656, Japan}
\affiliation{Center for Spintronics Research Network, The University of Tokyo,  Bunkyo-ku, Tokyo 113-8656, Japan}

\author{S. Ohya}
\affiliation{Department of Electrical Engineering and Information Systems, The University of Tokyo, Bunkyo-ku, Tokyo 113-8656, Japan}
\affiliation{Center for Spintronics Research Network, The University of Tokyo,  Bunkyo-ku, Tokyo 113-8656, Japan}

\author{A. Fujimori}
\affiliation{Department of Physics, The University of Tokyo, Bunkyo-ku, Tokyo 113-0033, Japan}


\date{\today}

\pacs{}

\maketitle


\section{Electronic configuration of F\lowercase{e} atom in the G\lowercase{e} host}
When an Fe atom substitutes for a Ge atom, the neutral Ge atom with the 3$d^{10}$($4sp$)$^{4}$ configuration is replaced by the neutral Fe atom with the 3$d^{6}$($4sp$)$^{2}$, 3$d^{5}$($4sp$)$^{3}$, or 3$d^{4}$($4sp$)$^{4}$ configuration. If the Fe atom takes the 3$d^{6}$($4sp$)$^{2}$ configuration (Fe$^{2+}$), two holes are provided to the Ge host since two 4$sp$ electrons are missing in the 3$d^{6}$($4sp$)$^{2}$ configuration compared to 3$d^{10}$($4sp$)$^{4}$.
An interstitial Fe$^{2+}$ atom acts as a double-donor, which donate the ($4sp$)$^2$ electrons to the Ge host.
Taking $\sim$15\% of interstitial Fe atoms \cite{Wakabayashi:2014ab} into account, the number of holes brought by one Fe atom can be estimated as $2\times0.85-2\times0.15=1.2$. 

\section{Resonance photoemission spectra in the angle-integrated mode}

\begin{figure}[h]
\begin{center}
\includegraphics[width=7 cm]{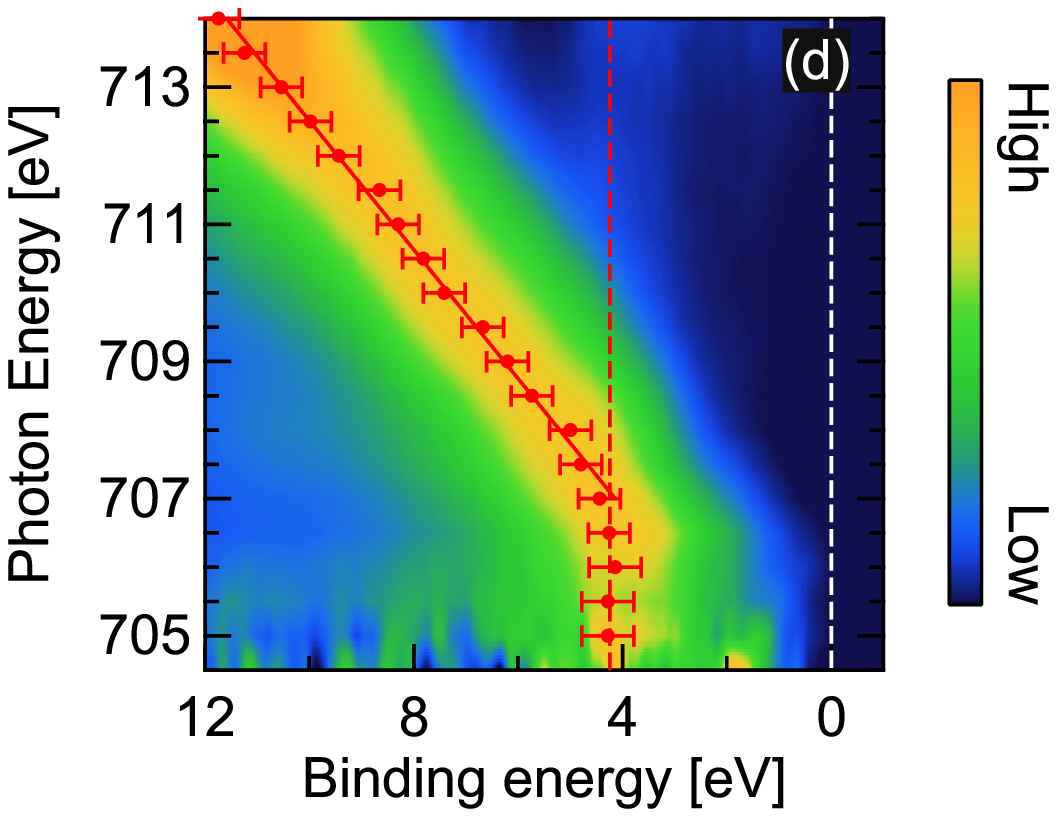}
\caption{False color plot of the spectra in Fig. 1(a) in the main text normalized to the area of the energy distribution curve (EDC). The solid circles represent the peak position of each spectrum. The dispersive Auger peak and the non-dispersive resonance feature are indicated by red solid and dashed lines, respectively. }
\label{colorPlot}
\end{center}
\end{figure}

Figure \ref{colorPlot} shows the false color plot of the spectra in Fig. 1(a) in the main text normalized to the area of the energy distribution curve (EDC). As is also stated in the main text, the dispersive normal Auger peak and non-dispersive feature around the binding energy of 4 eV can be seen, denoted by a red solid line and dashed line, respectively.
The latter structure has been found in various metallic compounds such as Ni \cite{Guillot:1977aa, Weinelt:1997aa}, Cr, Fe \cite{Hufner:2000aa}, and Fe-based superconductors \cite{Bondino:2008aa, Koitzsch:2010aa, Levy:2012aa}.
The energy of $4.2\pm0.2$ eV is larger than that of Fe metal (3.2 eV) \cite{Hufner:2000aa} but is close to that of the Fe-based superconductors CeFeAsO$_{0.89}$F$_{0.11}$ (4.2 eV) \cite{Bondino:2008aa} and BaFe$_{2}$As$_{2}$ (3.6eV) \cite{Koitzsch:2010aa}, which might be due to the similar tetrahedral coordination of non-metal atoms to the Fe atoms in the Fe-based superconductors and Ge:Fe.

\begin{figure}[h!]
\begin{center}
\includegraphics[width=7 cm]{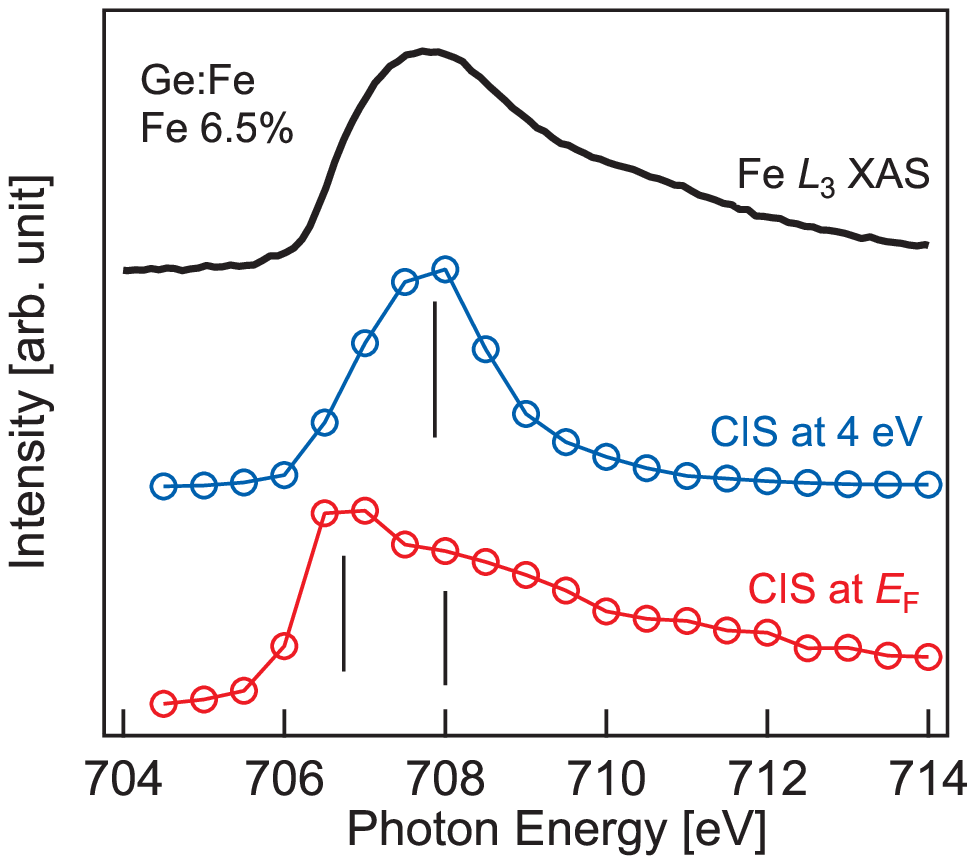}
\caption{Constant initial state (CIS) spectra around $E_{F}$ ($E_{B}=0.4$ to $-0.2$ eV) and at 4 eV. XAS spectra are also shown by a black solid curve.}
\label{FANO}
\end{center}
\end{figure}

Figure \ref{FANO} shows the intensity of the RPES spectrum around $E_{\rm F}$ (-0.2 to 0.4 eV) as a function of photon energy, namely, constant initial state (CIS) spectrum, where the XAS spectrum and the CIS spectrum at $E_{B}=$4 eV are also shown as references. 
The CIS spectrum show a sharp increase at the threshold well below the peak of the XAS spectrum, followed by a broader peak at a higher photon energy. The sharp increase may originate from the transitions into the narrow 3$d$($e$) band just above $E_{\rm F}$, while the broader peak from transitions into the empty 3$d$($t_{2}$) level, corresponding to the XAS spectrum.

\section{Resonance photoemission spectra in the Angle-resolved mode}

\begin{figure}[h]
\begin{center}
\includegraphics[width=11 cm]{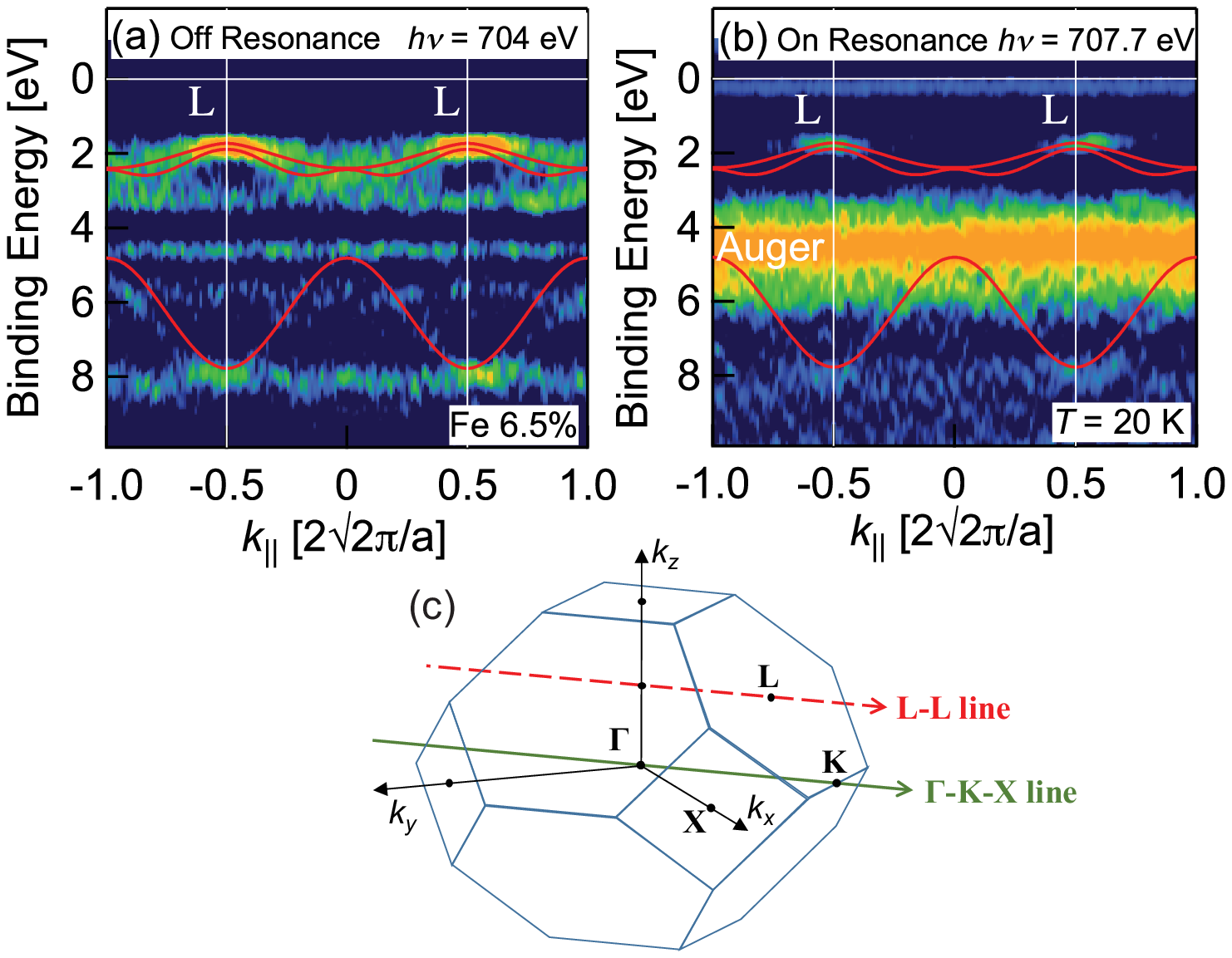}
\caption{(a), (b) EDC second derivatives of the ARPES spectra taken at (a) off-resonance ($h\nu$ = 704 eV) and (b) on-resonance (707.7 eV). Solid curve represents the calculated band dispersions of the Ge host along the L-L line in the Brillouin zone of the fcc lattice as shown by a red dashed arrow in panel (c). The $\Gamma$-K-X line is also shown by a green solid arrow.}
\label{Reso_ARPES}
\end{center}
\end{figure}

Figures \ref{Reso_ARPES}(a) and (b) shows the EDC second derivative of the photoemission spectra taken in the angle-resolved mode at $h\nu= 704$ eV (off-resonance) and 707.7 eV (on-resonance), respectively. These energies nearly correspond to the L-L line in the Brillouin zone of the fcc lattice (see Fig. \ref{Reso_ARPES}(c)) if the inner potential of $\sim$11 eV is assumed. How we estimated the value of 11 eV is described in the main text (see Figs. 2(a) and (b)). In the figure, calculated band dispersions of the Ge host are also shown by solid curves. 
Note that the flat structures seen in Fig. \ref{Reso_ARPES}(a) is not from Fe atoms, but just from the photoelectrons that lost the momentum information when escaping from the surface because those flat structures are not enhanced on resonance.
Apart from the intense Auger peak in the on-resonance spectra, the primary difference from the off-resonance spectra is the non-dispersive feature just below $E_{\rm F}$ as also discussed in the main text. This structure originates from the Fermi edge of the Fe 3$d$ states enhanced on resonance, as already confirmed by the angle-integrated spectra. This non-dispersive feature probably reflects the random distribution of the Fe atoms.

\bibliography{BibTex_all_GeFe.bib}